\documentclass[aps,preprint,nofootinbib,eqsecnum]{revtex4}
\usepackage{amsmath,amssymb,bm}
\usepackage{graphicx}
\usepackage{latexsym}
   \let\d=\partial

\let\w=\omega

\def\nn{\nonumber}
\def\imag{i}

\let\p=\partial
\def\be{\begin{equation}}
\def\ee{\end{equation}}
\def\bea{\begin{eqnarray}}
\def\eea{\end{eqnarray}}
\def\ba{\begin{array}}
\def\ea{\end{array}}

\def\ep{{\epsilon}}
\def\vep{{\varepsilon}}

\begin{document}
\title{Collective cyclotron motion of the relativistic plasma in graphene}
\author{Markus M\"uller}
\affiliation{Department of Physics,
Harvard University, Cambridge MA 02138, USA}
\author{Subir Sachdev}
\affiliation{Department of Physics,
Harvard University, Cambridge MA 02138, USA}
\date{\today}
\begin{abstract}
We present a theory of the finite temperature thermo-electric response functions
of graphene in the hydrodynamic regime where electron-electron collisions dominate the scattering.
In moderate magnetic fields, the Dirac particles undergo a collective
cyclotron motion with a temperature-dependent relativistic cyclotron frequency
proportional to the net charge density of the Dirac plasma. In contrast to the
undamped cyclotron pole in Galilean-invariant
systems (Kohn's theorem), here there is a finite damping induced by collisions
between the counter-propagating particles and holes.
This cyclotron motion shows up as a damped pole in the frequency dependent
conductivities, and should be readily detectable in microwave measurements at
room temperature. We also compute the large Nernst signal in the hydrodynamic regime which is significantly bigger than in ordinary metals.
\end{abstract}

\maketitle

\section{Introduction}

In the absence of gate voltage or external impurities,
graphene is a quantum critical system \cite{son,guinea,ys, gorbar,joerg}, whose quasiparticles
are governed by a
relativistic massless Dirac equation in two spatial
dimensions~\cite{Semenoff84,Haldane88,Zhou06}.
(By `quantum critical' we mean here a system in which
the inelastic scattering rate from electron-electron interactions is of order $k_B T /\hbar$ (up to logarithms),
and not models of localization transitions of free electrons~\cite{mirlin}.)
Experimental realizations of
relativistically-invariant systems are
rare, and so it is of interest to study physical phenomena which rely on the
Dirac nature of the electrons. In this paper we focus
our attention on the possibility of observing a relativistic cyclotron resonance
in the (collective) electronic motion in graphene.
This resonance shows up as a bump in the frequency dependence of all thermoelectric response functions.
As we will discuss in detail,
the cyclotron resonance frequency has clear signatures in its dependence upon
field ($B$), temperature ($T$), and electron density ($\rho$)
which distinguish it from the cyclotron resonance of non-relativistic electrons.
Moreover, the relativistic cyclotron resonance
is intrinsically damped by electron-electron interactions: this damping arises
from collisions between electrons and holes which
execute cyclotron orbits in opposite directions. In contrast, the damping of the
cyclotron resonance in metals arises primarily
from impurity scattering: Kohn's theorem~\cite{kohn} implies that electron-electron interactions do not broaden the cyclotron
resonance in Galilean-invariant systems. This theorem applies to metals with a single parabolic (non-relativistic) band, which is a reasonable approximate description for many simple Fermi surfaces. However, it fails for semimetals such as graphene both due to the linear, relativistic dispersion and the presence of two bands.

We shall be interested here in the cyclotron resonance in a hydrodynamic,
collision-dominated regime, where disorder plays only a minor role.
This regime exists at high temperature and is defined by the requirement that the vast majority of collisions arise from electron-electron interactions. This assures local equilibration before scattering from impurities occurs. The second requirement for hydrodynamics to apply is the following: The rate of deflection of linearly propagating, thermal quasiparticles due to the magnetic field must be small compared to the inelastic scattering rate, $\tau^{-1}_{\rm inel}$, arising from interactions. The same restriction must hold for the frequency of the external driving fields, $\omega\ll \tau^{-1}_{\rm inel}$. The quantum-critical nature of graphene \cite{son}
implies that the mean time between collisions is of order $\hbar/k_B T$. Since we
require that this inelastic collision time be shorter than the elastic scattering time from impurities, we cannot allow
$T$ to become too small, and will find that room temperature is suitable for
observing the physics we
are interested in. Further conditions controlling the range of parameters are
discussed below in Section~\ref{sec:conditions}.

The cyclotron resonance is formally defined as the pole of the thermo-electric response functions closest to the real axis in the complex frequency plane.
The expressions for the cyclotron frequency (its real part), $\omega_c$, and the damping
frequency (its imaginary part), $\gamma$, are
the same as those in Refs.~\onlinecite{nernst,hh}:
\begin{equation}
\label{omegac}
\omega_c \equiv \frac{e B \rho v_F^2}{c(\varepsilon + P)}~~~;~~~\gamma \equiv
\frac{\sigma_Q B^2 v_F^2}{c^2
(\varepsilon + P)}\,.
\end{equation}
Here $v_F$ is the `velocity of light' for the Dirac fermions, which was experimentally measured to be~\cite{Zhang05,stormer,Yacoby07} $v_F \approx 1.1\times 10^8 ~{\rm cm/s}\approx c/300$. The density of electrons $\rho$ is defined so that
$\rho=0$ for
undoped graphene. $\varepsilon$ and $P$
are the thermodynamic energy density and pressure of the Dirac plasma,
respectively, which are also measured with respect to undoped graphene. In the relativistic regime where the temperature exceeds the chemical potential $|\mu|$, the energy density and the pressure grow with the third power of temperature and assume typical values of
$\varepsilon, P\sim (k_B T)^3/(v_F\hbar)^2 = 3.28 \times 10^{12}\times T_{300}^3 {\rm meV}/{\rm cm}^2$
 where $T_{300}=T/300 {\rm K}$.
The coefficient $\sigma_Q$ will be seen to arise as a transport parameter in the hydrodynamic description of a relativistic fluid which cannot be fully determined by thermodynamics and hydrodynamics alone. It can, however, be computed in a microscopic approach, as recently carried out in Ref,~\cite{hydroBoltzmann}. The parameter $\sigma_Q$ has the units of an electrical conductivity and, as will become clear from later formulae, it describes the part of the d.c. conductivity which is independent of impurities, deriving solely from interactions, c.f., Eq.~(\ref{sigmaxx}).
At particle-hole symmetry, $\sigma_Q$ coincides with the finite conductivity of a clean system in zero field, $B=0$. In the relativistic, collision-dominated regime which is of foremost interest here, one has $\sigma_Q \sim \frac{1}{\alpha^2}\frac{e^2}{h}$~\cite{Lars} where $\alpha=\frac{e^2}{\epsilon_r\hbar v_F}\approx \frac{2.0}{\epsilon_r}$ is the fine structure constant of graphene characterizing the strength of Coulomb interactions, and $\epsilon_r$ is the dielectric constant due to the adjacent media. In general, $\sigma_Q$ it is a scaling function of $\mu/T$.
The dependence of the thermodynamic variables and the transport coefficient $\sigma_Q$ on temperature and chemical potential are further discussed in Section~\ref{sec:thermo}.

Physically, the cyclotron resonance is due to the tendency of fluid elements of the electron-hole plasma to undergo a circular motion at frequency $\omega_c$. This frequency results as an average over the left- and right-circulating orbital motions of thermally excited electrons and holes, which collide with each other many times before they would be able to complete a (non-interacting) cyclotron orbit. For this reason, $\omega_c$ is proportional to the excess charge density $\rho$, and vanishes at the particle-hole symmetric Dirac point where $\rho=0$. Nevertheless, a bump in the frequency dependent response around $\omega=0$ survives also in this case. Note that it broadens rather rapidly with decreasing $T$ as $\gamma\sim T^{-3}$.

In the non-relativistic regime $|\mu|\gg T$, where graphene turns into an ordinary Fermi liquid, the cyclotron resonance tends to the semiclassical value $\omega_c= e v_F^2 B/(c \mu)$ corresponding to the cyclotron frequency of non-interacting Dirac fermions at the chemical potential $\mu$. This should indeed be expected since all thermally excited quasiparticles share essentially the same non-interacting cyclotron frequency which is not altered by the interactions.

From microscopic transport theory one finds that $\sigma_Q$ decreases as $(T/\mu)^2$ in the non-relativistic Fermi liquid regime~\cite{hydroBoltzmann}. Accordingly, the intrinsic damping $\gamma$ decreases and the cyclotron resonance becomes increasingly sharp for small $T/\mu$.
We note that in addition to the intrinsic damping represented
by $\gamma$, there will also be extrinsic damping from impurity scattering, as is discussed
in Section~\ref{sec:conditions}. The complete expression for the frequency dependence of the conductivity
across the cyclotron resonance is given in Eq.~(\ref{sxxf}), and is illustrated in Fig.~\ref{sigmafig}.

Recent experiments \cite{stormer,deacon} have observed a ``non-hydrodynamic'' cyclotron resonance, in a regime
of strong magnetic fields in which the Landau levels can be resolved. As we will see below,
we are discussing here the different regime of weak fields and high temperatures, to which we hope the experiments will be extended.

\section{Model of graphene}
%Hamiltonian
We consider a single sheet of graphene described by the Hamiltonian
\begin{eqnarray}
\label{model}
H &=& H_0 + H_1 +H_\textrm{dis},
\end{eqnarray}
where the low energy tight binding part is
\begin{eqnarray}
H_0 &=&- \sum_{s}\sum_{V=K,K'}\int d \mathbf{x} \left[  \Psi_{s,V}^{\dagger} \left( iv_F \vec{\sigma}\cdot \vec{\nabla}
 + \mu\right) \Psi_{s,V} \right]\;,
\end{eqnarray}
with the Fermi velocity $v_F$, the two component wavefunction $\Psi_{s,V}$ describing the amplitude on the two sublattices for electrons with spin $s$ and momenta close to one of the two Fermi points $V=K, K'$.
A magnetic field is introduced as usual via minimal coupling.

The $1/r$ Coulomb interactions take the form
\begin{eqnarray}
&& H_1 = \frac{1}{2} \sum_{s,V} \sum_{s',V'} \int d \mathbf{x} d \mathbf{x'}
\Psi_{s,V}^{\dagger}(\mathbf{x})  \Psi_{s,V}(\mathbf{x})\frac{e^2}{\epsilon_r |{\mathbf{x}-\mathbf{x}'}|}
\Psi_{s',V'}^{\dagger}(\mathbf{x}')  \Psi_{s',V'}(\mathbf{x}').\nn
\end{eqnarray}
The term $H_\textrm{dis}$ describes the presence of weak disorder which induces elastic scattering and thus weak momentum relaxation at a rate $\tau$ proportional to the impurity concentration. In the high temperature regime which we are focusing on, the elastic scattering rate is smaller than the inelastic scattering rate and can be taken into account as a perturbation.

\section{The observability of the collective cyclotron resonance in graphene}
\label{sec:conditions}

As discussed in Ref.~\onlinecite{nernst}, the applicability of hydrodynamics to the system (\ref{model})
requires
that the magnetic field $B$ be weak, so that the Landau-quantization of thermal
excitations is not discernible, i.e.,
%:\bea
%\label{condition_on_B}
%(\hbar v)^2 \frac{2eB}{\hbar}\ll (k_BT)^2,
%\eea
\bea
\label{1stLL}
E_1\equiv \hbar v_F\sqrt{\frac{2eB}{\hbar c}}\ll k_BT,
\eea
where $E_1$ is the first Landau level for graphene at the Dirac
point~\cite{Haldane88,Novoselov05,Zhang05,Zhang06}.
This amounts to requiring that
\bea
B\ll B^*(T)\equiv \frac{\hbar c}{2e}\frac{(k_B T)^2}{(\hbar v_F)^2}= \left( 0.42\times
T_{300}^2\right)\, {\rm Tesla}.
\eea
For the hydrodynamic analysis to hold, one needs a slightly more stringent condition:
\bea
\label{Bcond}
E_1\ll \hbar \tau_{\rm inel}^{-1}\sim \alpha^2 k_B T,
\eea
where the last estimate applies to the relativistic regime $T<|\mu|$.
This condition expresses the requirement that the inelastic scattering rate, and thus the equilibration rate, should dominate the rate by which the magnetic field deflects electrons from their linear motion.
In the following we will assume $\alpha$ to be of order unity, but we nevertheless will retain $\alpha$ in most scaling estimates.

It will also be useful to express our electron densities in terms of
the characteristic density of thermal excitations,
\bea
\rho_{\rm th} \equiv \left(\frac{k_B T}{\hbar v_F}\right)^2=\left(1.27\times
10^{11}\times T_{300}^2\right)\, {\rm cm}^{-2}.
\eea
This density should be compared to disorder
induced density variations which are of order $\delta\rho_{\rm dis}\sim
10^{11} {\rm cm}^{-2}$~\cite{Yacoby07} varying on typical length scales $\sim 30 {\rm nm}$.
To remain close to quantum criticality, ensuring a universal conductivity due to thermal pair creation and annihilation processes, the regime $\delta\rho_{\rm dis}<\rho_{\rm
th}$ is preferred. This suggests measurements at room temperature or above.

As mentioned above, the cyclotron resonance occurs at
\bea
\omega_c&=& \frac{eB \rho v_F^2}{c(\varepsilon+P)}=\frac{1}{2\Phi_{\varepsilon+P}}\frac{k_B T}{\hbar} \frac{B}{B^*(T)}
\frac{\rho}{\rho_{\rm th}}\nn\\
&=& \frac{T_{300}}{\Phi_{\varepsilon+P}} \frac{B}{B^*(T)} \frac{\rho}{\rho_{\rm
th}}\, 1.96\times 10^{13}\, {\rm s}^{-1}, \label{omegaval}
\eea
where we have used a free-electron equation of state to determine the value of
$\varepsilon+P$, with
$\Phi_{\varepsilon+P}= (\varepsilon+P)(\hbar v)^2 T^{-3}$ being a dimensionless number
$\mathcal{O} (1)$, as given in Eq.~(\ref{eP(rho)}) below; we will also comment in Section~\ref{sec:thermo} on the effect of interactions on the equation of state.

The collective cyclotron frequency lies within the
hydrodynamic frequency regime if $\hbar\omega_c \ll \hbar\tau_{\rm inel}^{-1}\sim \alpha^2k_B T$ where the latter estimate applies to the relativistic, quantum-critical regime. This requires
\bea
\label{cond_hydro_rho}
\frac{\hbar\omega_c}{\alpha^2 k_B T}=\frac{1}{2\alpha^2
\Phi_{\varepsilon+P}}\,\frac{B}{B^*(T)}\frac{\rho}{\rho_{\rm th}} \ll 1.
\eea
For room temperature and values of $B$ and $\rho$ as suggested below, this falls
into the range of microwave frequencies. Thus the resonance should be readily detectable by
measuring the real and imaginary part of $\sigma_{xx}(\omega)$ of a graphene sample in a cavity.

For inelastic scattering to be the dominant relaxation process, i.e., one certainly
needs the impurity scattering time to satisfy $\tau\gg \tau_{\rm inel}$.
However, to observe the relativistic cyclotron resonance, one has to require the
more stringent condition
\bea
\label{tau}
\tau\gg 1/\omega_c = \left(\frac{k_B T}{\hbar \omega_c}\right) \,\frac{2.5\times
10^{-14}\,{\rm s}}{T_{300}}=
\frac{\Phi_{\varepsilon+P}}{T_{300}} \frac{B^*(T)}{B}\frac{\rho_{\rm th}}{\rho}
\,{5\times 10^{-14}\,{\rm s}},
\eea
to ensure that the disorder-induced broadening does not wash out the resonance.
Such long scattering times can indeed be achieved in high mobility graphene where
$\tau\sim 10^{-13}{\rm s}$ is a typical value.~\cite{Tan07,Yacoby07}

%According to Amir Yacoby the mean free time $\tau=\ell/v\approx 3\times 10^{-
%14}s$ ($\ell$ being estimated from the measured diffusion constant via
%$D=v\ell/2)$. However the purity can apparently be improved upon by 'cleaning'
%graphene with large currents.

Apart from a small impurity scattering one also needs the intrinsic broadening
of the cyclotron resonance due to electron-electron interactions to be smaller
than $\omega_c$:
\bea
\label{gammavswc}
\frac{\gamma}{\omega_c}=\frac{\Phi_{\sigma}}{4\pi}\,\frac{B}{B^*(T)}\frac{\rho_{
\rm th}}{\rho}< 1,
\eea
where $\Phi_\sigma=\sigma_Q/(e^2/h)= \mathcal{O} (1)$.

Note that (\ref{cond_hydro_rho}) and (\ref{gammavswc}) can be satisfied
simultaneously, e.g., with a choice of magnetic field $B/B^*(T)\sim
\mathcal{O}(0.1)$ and charge density ${\rho}/{\rho_{\rm th}}\sim \mathcal{O}(1)$.

%With this choice, and $T\sim 100 {\rm K}$, or even room temperature leads to
%cyclotron frequencies of order $\omega_c\sim 10^{11} {\rm Hz}$ at fields of
%order $B\sim 10 {\rm mT}$, which should be possible to detect in pure graphene.

\subsection{Scaling functions for thermodynamic variables}
\label{sec:thermo}

For the evaluation of thermodynamic state variables magnetic field effects can be safely
neglected since both Zeeman energy and Landau level splitting are significantly
smaller than $T$.
Further we will also neglect interactions and use the free theory to evaluate thermodynamic quantities.
This can be justified by the fact that the Coulomb interactions are
marginally irrelevant \cite{guinea,ys}. However, as emphasized by Son \cite{son}, the
bare value of the interactions are not small ($\alpha={\cal O}(1)$, and there may well be a significant regime
of intermediate scales where the interactions remain significant, and the theory is
characterized by a dynamic critical exponent $z \neq 1$. Over this intermediate regime, we
have $\varepsilon+P \sim T^{2+z}$. Below, we neglect such effects, and simply use the
free theory to obtain a numerical estimate of parameters.

\begin{figure}[htbp]
  \centering
  \includegraphics[width=4in]{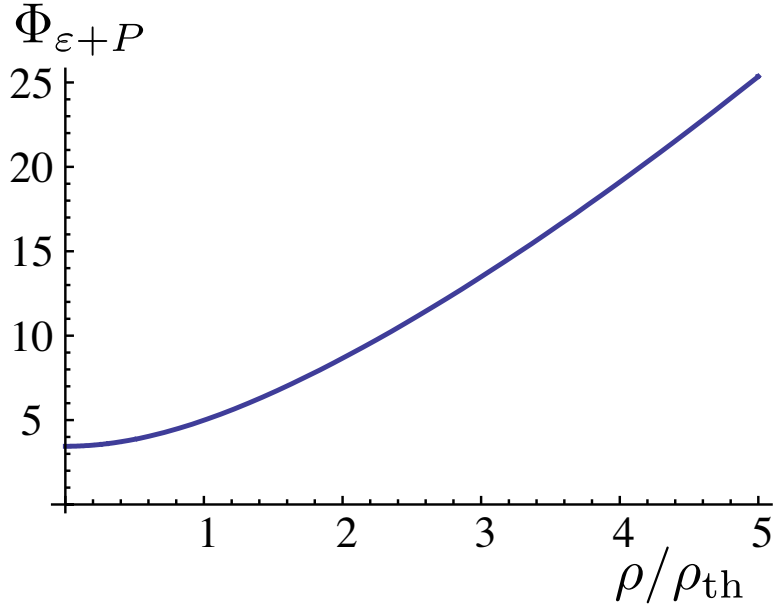}
  \caption{The dimensionless function $\Phi_{\varepsilon+P}$ of the density $\rho/\rho_{\rm th}$ for non-interacting
  Dirac fermions.}
  \label{energyfig}
\end{figure}

One obtains straightforwardly the scaling functions
\bea
\varepsilon+P= \frac{T^3}{(\hbar
v)^2}\Phi_{\varepsilon+P}(\tilde\mu=\mu/T)=T\rho_{\rm
th}\Phi_{\varepsilon+P}(\tilde\mu=\mu/T),
\eea
with
\bea
\Phi_{\varepsilon+P}(\tilde\mu)&=& 4\,\frac{3}{2}\int_0^\infty \frac{\tilde k\,
d\tilde k}{2\pi} \left[\frac{\tilde k}{e^{\tilde k-\tilde\mu}+1} + \frac{\tilde
k}{e^{\tilde k+\tilde\mu}+1}\right]\nn\\
&=& \frac{9\zeta(3)}{\pi } +
  \frac{6\,\log(2){\tilde \mu}^2}{\pi } +
  \frac{{\tilde \mu}^4}{8\,\pi } + \mathcal{O}({\tilde \mu}^6).
\eea
The prefactor $3/2$ derives from the relation $P=\varepsilon/2$, and the factor 4
accounts for spin and valley degeneracy. The electron density is
\bea
\rho= \frac{T^2}{(\hbar v)^2}\Phi_{\rho}(\tilde\mu=\mu/T)=\rho_{\rm
th}\Phi_{\rho}(\tilde\mu=\mu/T),
\eea
with
\bea
\Phi_{\rho}(\tilde\mu)&=&  4\int_0^\infty \frac{\tilde k\, d\tilde k}{2\pi}
\left[\frac{1}{e^{\tilde k-\tilde\mu}+1} - \frac{1}{e^{\tilde
k+\tilde\mu}+1}\right]\nn\\
&=& \frac{4\,\log (2)\,\tilde\mu}{\pi } +
  \frac{\tilde\mu^3}{6\,\pi } - \frac{\tilde\mu^5}{240\,\pi } + \mathcal{O}({\tilde
\mu}^7),
\eea
and thus
\bea
\label{eP(rho)}
\Phi_{\varepsilon+P}(\tilde\rho=\rho/\rho_{\rm th})&=&
\frac{9\zeta(3)}{\pi }
+\frac{3\pi \tilde\rho^2}{8\log (2)}
-\frac{3 {\pi}^3 \tilde\rho^4}
   {2048\,\log^4(2)}+\mathcal{O}(\tilde\rho^6)\nn\\
&=& 3.444 + 1.700\tilde\rho^2 -0.1968\tilde\rho^4 +\mathcal{O}(\tilde\rho^6).
\eea
A plot of this function is shown in Fig.~\ref{energyfig}
We note that it has been computed in the weak-coupling limit; if $\alpha={\cal O}(1)$ there will be order unity corrections from the Coulomb interactions which can be determined from the theory in Ref.~\onlinecite{son}.

For the results given in the next section the scaling function for the entropy
density $s=(\varepsilon+P-\mu\rho)/T$
will also be useful:
\bea
\frac{s}{\rho_{\rm th}}\equiv
\Phi_s(\tilde{\mu})=\Phi_{\varepsilon+P}(\tilde{\mu})-
\tilde\mu\Phi_{\rho}(\tilde{\mu}).
\eea

Finally we mention that the transport coefficient $\sigma_Q\equiv \frac{e^2}{h}\Phi_\sigma(\mu/T)$ is itself a scaling function of the ratio between chemical potential and temperature. In the relativistic regime, $T> \mu$, it can roughly approximated by a constant, whereas in the non-relativistic, Fermi liquid regime, $T<\mu$, a microscopic study shows that $\sigma_Q$ decreases as $(T/\mu)^2$. The precise scaling function has been obtained in Ref.~\onlinecite{hydroBoltzmann}.

\section{Derivation: Hydrodynamics of a relativistic fluid with Coulomb
interactions}
\label{sec:derive}

Here we discuss the magnetohydrodynamics of a relativistic fluid \cite{nernst} in the
presence of Coulomb interactions. Our general results for response functions in the hydrodynamic regime are then readily applied to graphene close to its Dirac point.
The characteristic velocity $v_F\neq c$ determines the relativistic dispersion, and the charge of (anti-)particles in the fluid is
$\pm e$. In the following we employ units with $v_F=e=\hbar=1$.

Because the Coulomb interactions spread with the speed of light $c\gg v_F$, they can be
considered instantaneous. They obviously break the relativistic invariance of
the fluid by singling out the lab frame as a particular reference frame, which we will eventually work below.

The stress energy tensor, $T^{\mu\nu}$, and the
current, $J^\mu$, of a relativistic fluid are expressed in terms of the three-velocity of the fluid element, $u^\mu\equiv (1,\vec{v})/\sqrt{1-v^2}$ ($u^\mu u_\mu = -1$, with $u^\mu = (1,0,0)$ in
the local rest frame for which the {\em energy} flow vanishes): ~\cite{LL,nernst}
\begin{eqnarray}
T^{\mu\nu} &=& (\vep+P) u^\mu u^\nu + P g^{\mu\nu} +
\tau^{\mu\nu},
\label{e0} \\
J^\mu &=& \rho u^\mu + \nu^\mu.
%\pa_\mu J^\mu_{(m)}&= & 0
\end{eqnarray}
These expressions include dissipative terms $\nu^\mu,\tau^{\mu\nu}$ which account for heat currents and viscous forces, respectively, and will be derived below.
Without the viscous terms, the stress energy tensor is a diagonal matrix in the rest frame of the fluid element with the pressure in the space-like entries,and energy density in the time-like entry.
In the lab frame (to lowest order in $v^i$) the components read
\bea
T^{00} &=& \vep, \\
T^{0i} &=& T^{i0} = (\vep+P)v^i, \\
T^{ij} &=& P \delta^{ij}+\tau^{ij}, \\
J^0 &=& \rho, \\
J^i &=& \rho v^i + \nu^i.
\eea
Here $\vep$, $P$, $\rho$, are functions of the local chemical
potential, $\mu(r)$, the local temperature $T(r)$ and the magnetic field $B$, as given
in the previous section for graphene.
Thereby the static Coulomb potential created by an inhomogeneous charge distribution is incorporated in $\mu(r)$.

%For simplicity we neglect the viscous forces, setting $\tau^{ij}=0$.
%The vector $\vec \nu$ capturing the dissipative current due to heat flow, and
%the tensor $\tau^{ij}$ describing the viscous contribution to the stress-energy
%tensor will be derived below.

%The above expressions for stress energy tensor and current hold a priori for fluids in the absence of long range forces.
%One then
%has to be careful when identifying the effect of an external perturbation, such
%as $\delta T$ or $\vec E$, on $\mu$.

The conservation laws for charge, energy and momentum read:
\begin{eqnarray}
%\pa_\beta \overline{J}^{\beta} &=& 0,
%\pa_\beta \overline{T}_{\alpha}^{\beta}= F_{\alpha\gamma}\overline{J}^\gamma.
\p_\beta {J}^{\beta} &=& 0,\\
\p_\beta {T}^{\beta\alpha}&=& F^{\alpha\gamma}{J}_\gamma,
\end{eqnarray}
where the electromagnetic field tensor,
\begin{equation}
F^{\mu \nu} = \left( \begin{array}{ccc} 0 & E_x & E_y \\
-E_x & 0 & B \\
-E_y & -B & 0
\end{array} \right), \label{fmn}
\end{equation}
contains a self-generated, spatially varying electric field due to the
inhomogeneous charge density of the system itself:
\bea
\vec{E}&=&-\vec\nabla \phi,\nn\\
\label{phi}
\phi(x)&=& \int d^2 y\, U(x-y)\,\left[\rho(y)-\rho\right],
\eea
with
\bea
U(r) &=& \frac{e^2}{r} = \int \frac{d^2k}{(2\pi)^2}\, U_k \exp[i \vec k\cdot \vec
r]\quad\quad;\quad\quad U_k = \frac{2\pi \alpha}{|k|},\nn\\
\vec{E}(\vec k)&=&-i\vec k U_k \rho(\vec k),\label{Ek}
\eea
where the uniform background charge density $\rho$ has been subtracted in
(\ref{phi}).

In the lab frame the linearized conservation laws read  more explicitly:
\bea
\p_t \rho+\vec\nabla\cdot \vec J&=&0, \label{chargecons}\\
\p_t \varepsilon +\vec\nabla \cdot[(\vep+P)\vec v]&=&\rho\vec v\cdot\vec E ,
\label{econs}\\
\left(\p_t+\tau^{-1}\right) (\varepsilon+P) \vec v+\vec \nabla P&=&B \hat
\epsilon\vec J+ \rho\vec{E}, \label{momcons}
\eea
where in addition we have included a momentum relaxation time $\tau$ due to
weak impurity scattering. Note that the latter also breaks the relativistic invariance, and accordingly the relativistic hydrodynamics should only be expected to hold as long as $\tau^{-1}\ll \tau^{-1}_{\rm inel}$.
The above set of equations is closed by the constitutive equation for the current
\bea
\vec{J} &=& \rho \vec v+\vec \nu.
\eea

\subsection{Heat current and viscous terms in a magnetic field}
With the help of the thermodynamic relations
\bea
P+\varepsilon=s T +\mu \rho,\quad d\varepsilon=T ds+\mu d\rho,
\eea
and employing the conditions~\cite{LL} $u_\mu\nu^\mu =u_\mu\tau^{\mu\nu}=0$,
the energy conservation law ($u_\mu\d_\nu T^{\mu\nu}=u_\mu F^{\mu\nu}J_\nu$) can
be rewritten in the form of a law of entropy production,
\begin{equation}
\label{entropyprod}
\d_\mu \left(su^\mu - \frac\mu T \nu^\mu\right) =
   -\nu^\mu \d_\mu \left(\frac\mu T\right) + \frac{1}{T} F_{\mu\nu} u^\nu
\nu^\mu - \frac{\tau^{\mu\nu}}
T \d_\mu u_\nu\,.
\end{equation}
It is natural to interpret the left hand side as the divergence of an entropy
current,
\bea
S^\mu = su^\mu - \frac\mu T \nu^\mu,
\eea
which, by the second law of thermodynamics, has to be positive.
For small velocity derivatives $\d_\mu u^\nu$ and electromagnetic fields $B, E$,
the dissipative terms should be linear in these perturbations, and hence one
deduces the form
\begin{align}
\nu^\alpha&=-\sigma_Q\left[ T(g^{\alpha\lambda}+u^\alpha
u^\lambda)\partial^\alpha(\mu/T)-F^{\alpha\lambda}u_\lambda\right]
,\label{nu-constit}\\
   \tau^{\mu\nu} &= -\left(g^{\mu\lambda}+ u^\mu u^\lambda\right)\left[ \eta
(\d_\lambda u^\nu + \d^\nu u_{d\lambda})
   +\left(\zeta-\eta\right)
    \delta^\nu_\lambda \d_\alpha u^\alpha\right]\,,
\end{align}
where $\sigma_Q$ is a conductivity (of order $e^2/h$), and $\eta$ and $\zeta$
are the shear and bulk viscosities, as will be clear from the expression (\ref{viscosity}) given
below. For graphene at the Dirac point, $\eta, \zeta \sim T^2/\alpha^2$. These can be
computed by solving a linearized kinetic Boltzmann
equation~\cite{ssqhe,Lars}.

The relativistic expression (\ref{nu-constit})
%\bea
%\nu^\alpha=-\sigma_Q\left[ T(g^{\alpha\lambda}+u^\alpha
%u^\lambda)\partial^\alpha(\mu/T)-F^{\alpha\lambda}u_\lambda\right]
%\eea
reduces to a spatial vector in the lab frame %$u^\mu=(1,0,0)$:
% where $\nu^0=u^\mu\nu_\mu=0$, which is required for $\rho$ to denote the
%particle density:
\bea
\vec{\nu} &= -\sigma_Q \left( \vec\nabla \mu -\frac{\nabla T}{T}\mu -
B\hat\epsilon\vec v - \vec E\right),
\eea
%The viscous contribution to the stress energy tensor must take the form
%\bea
%\tau^{\mu\nu} &= -(g^{\mu\lambda}+ u^\mu u^\lambda)\left[ \eta (\d_\lambda
%u^\nu + \d^\nu u_{\lambda})
%   +\left(\zeta-\eta\right)
%    \delta^\nu_\lambda \d_\alpha u^\alpha\right]\,,
%\eea
%where $\eta$ and $\zeta$ are the shear and bulk viscosities, as is clear from their form
while the viscous contribution to the stress energy tensor takes the familiar
form
\bea
\label{viscosity}
\tau^{ij} &= \left[ \eta (\d_i u^j + \d_j u^i)
   +\left(\zeta-\eta\right)\delta^{ij}\, \vec\nabla\cdot \vec v \right].
\eea

\subsection{Linear thermoelectric response}
The thermoelectric transport coefficients describing the current ($\vec J$) and
heat current ($\vec Q$) response to electric fields and temperature gradients
are defined by the relation
\begin{equation}
\left( \begin{array}{c} \vec{J} \\ \vec{Q} \end{array} \right) =
\left( \begin{array}{cc} \hat{\sigma} & \hat{\alpha} \\  T \hat{\tilde \alpha} &
\hat{\overline{\kappa}} \end{array} \right)
\left( \begin{array}{c} \vec{E} \\ -\vec{\nabla} T \end{array} \right),
\label{alltrans}
\end{equation}
where $\hat{\sigma}$, $\hat{\alpha}$, $\hat{\tilde \alpha}$ and
$\hat{\overline{\kappa}}$ are $2 \times 2$
matrices acting on the spatial indices $x,y$.
Rotational invariance in the plane imposes the form
\bea
\hat{\sigma}=\sigma_{xx}\, \hat{1} +\sigma_{xy}\hat{\epsilon},
\eea
where $\hat{1}$ is the identity, and $\hat{\epsilon}$ is the antisymmetric
tensor $\hat{\epsilon}_{xy}=-\hat{\epsilon}_{yx}=1$. $\sigma_{xx}$ and
$\sigma_{xy}$ describe the longitudinal and Hall conductivity, respectively.
An analogous form holds for the thermoelectric
conductivities $\hat{\alpha}, \hat{\tilde\alpha}$ which determine the Peltier,
Seebeck, and Nernst
effects, as well as for the matrix $\hat{\overline{\kappa}}$ which governs
thermal transport in the absence of electric fields. The latter applies to
samples connected to conducting leads, allowing for a stationary current flow.
In contrast, the thermal conductivity, $\hat{\kappa}$, is defined as the heat
current
response to $-\vec{\nabla} T$ in the absence of an electric
current (electrically isolated boundaries), and is given by
\begin{equation}
\hat{\kappa} = \hat{\overline{\kappa}} -T \hat{\tilde\alpha} \hat{\sigma}^{-1}
\hat{\alpha}. \label{kappadef}
\end{equation}

\subsection{Response functions from hydrodynamics}
We will now use the conservation laws (\ref{chargecons},\ref{econs},\ref{momcons})
to solve for the slow relaxation dynamics towards equilibrium, starting from a small initial long-wavelength perturbation. From the full solution of the relaxation dynamics in linear response approximation one can then determine the thermo-electric response functions in the hydrodynamic regime.

As dynamic variables we choose $T$, $\mu$ (which includes the static Coulomb potential), and
$v^x$ and $v^y$, and write
\begin{eqnarray}
\mu (r, t) &=& \mu + \delta \mu (r,t),
\nonumber \\
T (r, t) &=& T + \delta T (r,t). \label{pert}
\end{eqnarray}

The other variables,  $\varepsilon$, $P$, and $\rho$ are constrained by
local thermodynamic equilibrium to have the form
\begin{eqnarray}
\rho (r,t) = \rho + \delta \rho &\equiv & \rho +
\left. \frac{\partial \rho}{\partial \mu}  \right|_{T} \delta \mu + \left.
\frac{\partial \rho}{\partial T}  \right|_{\mu} \delta T,
\nonumber \\
\varepsilon (r,t) = \varepsilon + \delta \varepsilon &\equiv& \varepsilon +
\left. \frac{\partial \varepsilon}{\partial \mu} \right|_{T} \delta \mu
+ \left. \frac{\partial \varepsilon}{\partial T} \right|_{\mu} \delta T,
\nonumber \\
P (r,t) = P + \delta P &\equiv& P +  \rho \delta \mu + s \delta T,
\end{eqnarray}
where $\delta\mu$ is the deviation of the total electrochemical potential form the equilibrium value $\mu$, and
$\p \rho/\p \mu|_T \equiv \chi$ is the susceptibility. Note, that even though for
graphene at the Dirac point ($\mu=0$) $\chi(k)\sim k\to 0$ at $T=0$, the
susceptibility is always finite, $\chi\sim T$, for the hydrodynamic regime considered here ($T>0$).

Following the technique of Kadanoff-Martin~\cite{km} to derive hydrodynamic
response functions, we prepare the system in a state of local equilibrium as
characterized by slowly varying initial conditions $T(\vec r)$ and $\mu(\vec r)$, as created by an external electric field.
Local equilibrium implies that the screening of an inhomogeneous charge density must be selfconsistently built into the initial conditions. The relation between the actual
initial variation of the chemical potential, $\delta \mu^0$, and both an applied external
field $\vec E^{\rm ext}\equiv -\vec \nabla\left(\delta\mu^{\rm ext}\right)$ and temperature deviations $\delta T^0$, is therefore not entirely trivial: The total chemical potential $\mu(r)$ is the sum of the externally applied potential $\delta\mu^{\rm ext}$, and the Coulomb potential created by the total induced charge density.
In $k$-space, the Thomas-Fermi (RPA) screened Coulomb potential induced by the two perturbations leads to the chemical potential
\bea
\delta\mu^0(k)=\delta\mu^{\rm ext}(k)- \frac{U_k}{1+U_k \left. \partial \rho/\partial
\mu \right|_{T}} \left(\left. \frac{\partial \rho}{\partial \mu}  \right|_{T}
\delta \mu^{\rm ext}(k) + \left. \frac{\partial \rho}{\partial T} \right|_{\mu}
\delta T^0(k)\right).
\eea

After a Fourier transform in space and a Laplace transform in time,
Eqs.~(\ref{chargecons}-\ref{momcons}) together with (\ref{Ek}) take the form
\begin{eqnarray}
\label{mhd}
&&\w \left(\left.\frac{\partial \vep}{\partial \mu}\right|_{T} \delta \mu  +
\left.\frac{\partial \vep}{\partial T} \right|_{\mu} \delta T\right) - k
(\ep+P)v_{\parallel} \nn\\
&&\quad\quad\quad\quad =i \left[ \left.\frac{\partial \vep}{\partial \mu}
\right|_{T} \delta\mu^0 + \left.\frac{\partial \vep}{\partial T}
\right|_{\mu} \delta T^0\right] \,,  \\
&&\w \left(\left.\frac{\partial \rho}{\partial \mu} \right|_{T} \delta \mu +
\left.\frac{\partial \rho}{\partial T} \right|_{\mu} \delta T\right) - k
\left(\rho v_\parallel +
\sigma_Q \left[B  v_{\perp}+E_k\right] \right)
+ i \sigma_Q k^2  \left( \delta \mu - \frac{\mu}{T} \delta T\right) \nn\\
&&\quad\quad\quad\quad = i \left[\left.\frac{\partial \rho}{\partial \mu}
\right|_{T} \delta\mu^0 + \left.\frac{\partial \rho}{\partial T}
\right|_{\mu} \delta T^0\right] \,,  \nn\\
&&\left(\w+\frac{i}{\tau}\right) (\ep + P) v_{\parallel} - k (\rho \delta \mu +
s \delta T) - i
\rho B v_\perp +
i \sigma_Q B^2  v_{\parallel}  -i\rho E_k+ i k^2 (\eta + \zeta) v_{\parallel}
 =  i(\vep + P) v_{\parallel}^0 \,, \nn\\
&&\left(\w+\frac{i}{\tau}\right) (\ep + P) v_{\perp} + k \sigma_Q B  \left(
\delta \mu -
\frac{\mu}{T} \delta T\right) + i \rho B v_\parallel + i \sigma_Q B\left[
B v_\perp +E_k\right]+ i k^2 \eta v_\perp
 =  i(\vep + P )
v_\perp^0  \,,\nn
\end{eqnarray}
where
\bea
E_k=-i k U_k \left(\left.
\frac{\partial \rho}{\partial \mu}  \right|_{T} \delta \mu + \left.
\frac{\partial \rho}{\partial T}  \right|_{\mu} \delta T\right)\,.
\eea
In these expressions, $\delta\mu^0$, $\delta T^0$, $\delta
v_\parallel^0$ and $\delta v_{\perp}^0$ are the initial values (depending on
the wavevector $k$), while $\delta\mu$, $\delta T$, $\delta
v_\parallel$ and $\delta v_{\perp}$ are functions of $k$ and $\omega$.
The projections of $\vec v$ parallel and orthogonal to $\vec k$ are
$v_\parallel=\vec v\cdot \vec k/k$, $v_\perp= \vec k/k\cdot\hat \epsilon\vec v$.

%Further
%\bea
%\chi_k=\int d^2 x \frac{\partial \rho(0)}{\partial \mu(x)} |_{T}\exp[i \vec
%k\cdot \vec x].
%\eea

As shown in detail in Refs.~\cite{km,nernst}, the linear response functions can
be obtained in full generality and in closed form from the solution of the above
equations, e.g., $\sigma_{xx}(\omega,k)= \omega J_\parallel(\omega,k)/\delta
\mu^0$, where $J_\parallel=\vec J\cdot \vec k/k$.

We have restricted ourselves to the lowest non-trivial order in an expansion in
$k/\omega$, where Coulomb effects become visible. This will be sufficient to
describe the response to external perturbations of electromagnetic origin (e.g.,
microwaves) for which one always has $ck/\omega<\sqrt{\epsilon_r}$. When applying the
relativistic hydrodynamics to graphene with a characteristic velocity
$v/c\approx 1/300$, this implies the relation $vk/\omega <(v/c)
\sqrt{\epsilon_r}\ll 1$, which justifies the expansion to lowest order in
$k/\omega$.
The full $k,\omega$-dependence can be obtained in closed form. However, it is very
involved and does not contain much more physical information, so we do not report it
explicitly here.
%The dependence on larger $k$ may perhaps be probed by neutron scattering
%measurements.

\section{Results for the thermo-electric response}
\subsection{Limit of vanishing magnetic field}
In the limit of vanishing magnetic field $B\to 0$, the transverse response
vanishes for symmetry reasons. The longitudinal transport coefficients take the
relatively simple forms
\bea
\label{sigmaxx}
\sigma_{xx}(\omega,k;B=0)&=&\left(\sigma_Q + \frac{\rho^2}{P +
\varepsilon}\frac{\tau}{1-i\omega\tau}\right) \left[1-2\pi\alpha\frac{i k}{\omega}
\left(\sigma_Q +\frac{\tau}{1-i\omega\tau} \frac{\rho^2}{P + \varepsilon}\right)
\right]+ {\cal O}(k^2),
\\
\alpha_{xx}(\omega,k;B=0)&=& \tilde \alpha_{xx}(\omega,k;B=0) + {\cal O}(k^2)\\
 &=& \left(-\sigma_Q\frac{\mu}{T} +\frac{s\rho}{P+\varepsilon} \frac{\tau}{1-
i\omega \tau}\right) \left[1-2\pi\alpha\frac{i k}{\omega} \left(\sigma_Q +\frac{\tau}{1-
i\omega\tau} \frac{\rho^2}{P + \varepsilon}\right) \right] +
  {\cal O}(k^2), \nn\\
%&&+\frac{g}{T}  \frac{\,\left( sT\,\rho \,\tau   - \mu \,{{\sigma }_Q}
%(P+\epsilon)(1 - \imag \tau \,\omega)  \right) \,\left( \imag \,{\rho }^2\,\tau
%+   \left( P + \epsilon  \right) \,\left( \imag  + \tau \,\omega  \right)
%\,{{\sigma }_Q} \right) \,k}{    {\left( P + \epsilon  \right) }^2\,\omega
%\,{\left( \imag  + \tau \,\omega  \right) }^2} +  {\cal O}(k^2)\nn\\
\overline{\kappa}_{xx}(\omega,k;B=0)&=&
\left({\sigma }_Q\frac{{\mu }^2}{T}+\frac{s^2\,T}{P + \varepsilon}\frac{\tau}{1
-i\omega\tau} \right) \left[1-2\pi\alpha\frac{i k}{\omega} \left(\sigma_Q
+\frac{\tau}{1-i\omega\tau} \frac{\rho^2}{P + \varepsilon}\right) \right]\nn\\
  && \quad +2\pi \alpha i\frac{k}{\omega}  \sigma_Q \frac{P + \varepsilon}{T}
    \frac{\tau}{1  - i\omega\tau} +
  {\cal O}(k^2),\\
  \label{kappaxx}
\kappa_{xx}(\omega,k;B=0)&=&
\sigma_Q\frac{(P+\varepsilon)^2}{T}\frac{1}{\rho^2+(\sigma_Q/\tau)(P+\varepsilon) (1 - i
\omega\tau)} +
  {\cal O}(k^2).
\eea

Note that all response functions contain a piece proportional
to $ \sigma_Q$ which is independent of the impurity scattering time $\tau$ and thus is solely governed by the universal Coulomb interactions.
A second term proportional to $\tau/(1-i\omega \tau)$ has the form of a
classical Drude-like term which is due to the slow relaxation of the "momentum mode"~\cite{Lars}. This is an excitation of the electron-hole liquid which cannot relax due to Coulomb interactions because of their translational invariance. As one may expect for weak impurity concentration, the term contributed by the momentum mode grows linearly with the impurity scattering time $\tau$.
%In the nonrelativistic limit where
%$\rho/(\varepsilon+P)\to 1/m$ takes the role of an effective mass.

Note the simple structure of the leading $k$ dependence of the response functions: It only depends on Coulomb interactions (via $\alpha$), while the viscosities
$\eta, \zeta$ do not appear at this order. Including finite $k$ introduces a simple
factor
\bea
1-2\pi\alpha\frac{i k}{\omega} \left(\sigma_Q +\frac{\tau}{1-i\omega\tau}
\frac{\rho^2}{P + \varepsilon}\right) =1- 2\pi\alpha\frac{i k}{\omega}
\sigma_{xx}(\omega,k=0),
\eea
which multiplies the $k\to 0$ result. Interestingly, the thermal conductivity $\kappa$
(in the absence of currents) is not affected by Coulomb interactions to lowest
order.

\subsection{Magneto-transport}
\begin{figure}[htbp]
  \centering
  \includegraphics[width=4in]{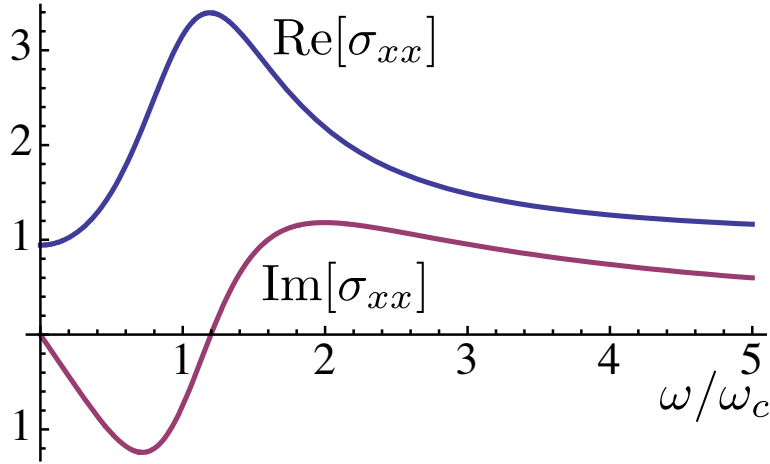}
  \caption{The real imaginary and imaginary parts of $\sigma_{xx}$, in units of $\sigma_Q$,
  for $\gamma/\omega_c = 0.3$ and $\omega_c \tau = 3$.}
  \label{sigmafig}
\end{figure}

Most interesting for our purpose is the response in a weak magnetic field for which we obtain:
\bea
\sigma_{xx}(\omega,k)&=& {\sigma }_Q \frac{\left(\omega + \imag/\tau \right)
\left( \omega + i/\tau + i\gamma  + i \omega_c^2/\gamma  \right) }{
\left(\omega+ i/\tau + i\gamma \right)^2 -\omega_c^2  } \label{sxxf} \\
&&\quad\times
\left[1- 2\pi\alpha\frac{i k}{\omega} {\sigma }_Q \frac{\left(\omega + \imag/\tau
\right)
\left( \omega + i/\tau + i\gamma  + i \omega_c^2/\gamma  \right) }{
\left(\omega+ i/\tau + i\gamma \right)^2 -\omega_c^2  }\right] + {\cal
O}(k^2),\nn\\
\sigma_{xy}(\omega,k)&=& -\frac{\rho}{B}
\frac{\omega_c^2+\gamma^2+2\gamma(1/\tau-i\omega)}
{ \left(\omega+ i/\tau + i\gamma \right)^2 -\omega_c^2} \label{sxyf} \\
&&\quad\times \left[1- 2\pi\alpha \frac{i k}{\omega} {\sigma }_Q \frac{\left(\omega +
\imag/\tau \right)
\left( \omega + i/\tau + i\gamma  + i \omega_c^2/\gamma  \right) }{
\left(\omega+ i/\tau + i\gamma \right)^2 -\omega_c^2  }\right] + {\cal
O}(k^2).\nn
\end{eqnarray}
We plot the frequency dependence of the longitudinal conductivity in Fig~\ref{sigmafig}.
The collective cyclotron frequency $\omega_c$ and the intrinsic,
interaction induced damping frequency $\gamma$ were given in Eq.~(\ref{omegac}).

%&&+
%\frac{gk}{\omega}  \frac{\tau \,\left( \imag  + \tau \,\omega  \right)
%\,{{{\sigma %}_Q}}^2\,{{\omega }_c}\,
 %    \left(\left( 1 + \gamma \,\tau  - \imag \,\tau \,\omega  \right)  + \tau
%\,{{{\omega }_c}}^2/\gamma \right) \,\left(\left( 2 + \gamma \,\tau  -  2\imag
%\,\tau \,\omega  \right) + \tau \,{{{\omega }_c}}^2/\gamma \right)}{ {\left(
%{\left( 1 + \gamma \,\tau  - \imag \,\tau \,\omega  \right) }^2 + {\tau
%}^2\,{{{\omega }_c}}^2       \right) }^2} + {\cal O}(k^2)\nn\\
For the thermoelectric response functions, we have
\begin{eqnarray}
\tilde\alpha_{xx}(\omega,k) &=& -\frac{(\omega+i/\tau)\left( \sigma_Q
(\mu/T)(\omega+i/\tau+i\gamma) -i s  \rho/(\varepsilon+P)\right)}{ \left(\omega+
i/\tau + i\gamma \right)^2 -\omega_c^2  }\nn\\
&&\quad\times
\left[1- 2\pi\alpha\frac{i k}{\omega} {\sigma }_Q \frac{\left(\omega + \imag/\tau
\right)
\left( \omega + i/\tau + i\gamma  + i \omega_c^2/\gamma  \right) }{
\left(\omega+ i/\tau + i\gamma \right)^2 -\omega_c^2  } \right]+ {\cal
O}(k^2),\nn\\
&=& i\frac{(\omega+i/\tau)\left( s\omega_c/B +i\sigma_Q
(\mu/T)(\omega+i/\tau+i\gamma)\right)}{ \left(\omega+ i/\tau + i\gamma \right)^2
-\omega_c^2  }\\
&&\quad\times
\left[1- 2\pi\alpha\frac{i k}{\omega} {\sigma }_Q \frac{\left(\omega + \imag/\tau
\right)
\left( \omega + i/\tau + i\gamma  + i \omega_c^2/\gamma  \right) }{
\left(\omega+ i/\tau + i\gamma \right)^2 -\omega_c^2  }\right]+ {\cal
O}(k^2),\nn\\
%\frac{\left( \imag  + \tau \,\omega  \right) \,    \left( -\imag \,\rho \,\tau
%+
%       \mu \,{{\sigma }_Q}\,\left(  \left( \imag  + \imag \,\gamma \,\tau  +
%\tau \,\omega  \right)  +
%          \imag \,\tau \,{{{\omega }_c}}^2/\gamma \right)  \right) }{ {\left( 1
%+ \gamma \,\tau  - \imag \,\tau \,\omega  \right) }^2 + {\tau }^2\,{{{\omega
%}_c}}^2 } \\
%       &&- \frac{g k}{\omega}\frac{\imag \,{\left( \imag  + \tau \,\omega
%\right) }^2\,{{\sigma }_Q}\,
%     \left(\left( 1 + \gamma \,\tau  - \imag \,\tau \,\omega  \right)  + \tau
%\,{{{\omega }_c}}^2/\gamma
%       \right) \,\left( - \rho \,\tau  +
%       \mu \,{{\sigma }_Q}\,\left(\left( 1 + \gamma \,\tau  - \imag \,\tau
%\,\omega  \right)  +
%          \tau \,{{{\omega }_c}}^2/\gamma \right)  \right) }{{\left( {\left( 1
%+ \gamma \,\tau  - \imag \,\tau \,\omega  \right) }^2 + {\tau }^2\,{{{\omega
%}_c}}^2
%         \right) }^2}\nn\\
\alpha_{xx}(\omega,k) &=&\tilde\alpha_{xx}(\omega,k) + {\cal O}(k^2),\\
\tilde\alpha_{xy}(\omega,k) &=& -\frac{s}{B}
  \frac{\omega_c^2+\gamma^2 -i\gamma \left( \omega+ \imag/\tau \right) [1-
\mu\rho/(s T)]}{ \left(\omega+ i/\tau + i\gamma \right)^2 -\omega_c^2}\\
&&\quad\times
\left[1- 2\pi\alpha \frac{i k}{\omega} {\sigma }_Q \frac{\left(\omega + \imag/\tau
\right)
\left( \omega + i/\tau + i\gamma  + i \omega_c^2/\gamma  \right) }{
\left(\omega+ i/\tau + i\gamma \right)^2 -\omega_c^2  }\right] + {\cal
O}(k^2),\nn\\
\alpha_{xy}(\omega,k) &=&\tilde\alpha_{xy}(\omega,k) -
2\pi\alpha \frac{k}{\omega} \frac{B \sigma_Q^2}{T} \frac{\left( \omega +i/\tau\right) }{
{\left(\omega+ \imag/\tau  + \imag \,\gamma\right) }^2 -
    {{{\omega }_c}}^2} +{\cal O}(k^2).
%&&+\frac{g k}{\omega}\frac{\,\tau \,\left( \imag  + \tau \,\omega  \right)
%\,{{\sigma }_Q}\,
%     \left( \gamma \,\left( -\left( \mu \,\rho \,\left( 1 - \imag \,\tau
%\,\omega  \right)  \right)  +
%          s\,T\,\left( 1 + \gamma \,\tau  - \imag \,\tau \,\omega  \right)
%\right)  +
%       s\,T\,\tau \,{{{\omega }_c}}^2 \right) \,
%     \left( 1 + \gamma \,\tau  - \imag \,\tau \,\omega  + \frac{\tau
%\,{{{\omega }_c}}^2}{\gamma } \right) }
%     {B \,{\left( {\left( 1 + \gamma \,\tau  - \imag \,\tau \,\omega  \right)
%}^2 +
%         {\tau }^2\,{{{\omega }_c}}^2 \right) }^2}
%\\
\eea
Notice that the Onsager reciprocity $\alpha=\tilde\alpha$ only holds for the
response to homogeneous perturbations (i.e., $k=0$), while there are deviations
at finite $k$.

Finally, for the thermal conductivities, we have
\bea
\overline{\kappa}_{xx}(\omega,k) &=&
\frac{-\gamma \frac{P + \varepsilon}{T} -
    i \frac{s^2 T}{P + \varepsilon}\, \left( \omega +i/\tau \right)   +
     \sigma_Q \frac{\mu^2}{T}\,\left( \omega+ i/\tau\right) \,
     \left(\omega+ i/\tau + i\gamma\right) }{\left( \omega+i/\tau +
i\gamma\right)^2 -\omega_c^2}\nn \\
  &&\quad \times  \left[1- 2\pi\alpha\frac{i k}{\omega} {\sigma }_Q \frac{\left(\omega +
\imag/\tau \right)
\left( \omega + i/\tau + i\gamma  + i \omega_c^2/\gamma  \right) }{
\left(\omega+ i/\tau + i\gamma \right)^2 -\omega_c^2  }\right]\nn\\
&& \quad -2\pi\alpha\frac{k}{\omega}  \sigma_Q\frac{P +
\varepsilon}{T}\frac{\omega+i/\tau}{  \left( \omega+i/\tau  + i\gamma  \right)^2
- \omega_c^2}  +
{\cal O}(k^2),
\\
\kappa_{xx}(\omega,k) &=&  i\frac{(\varepsilon+P)}{T} \frac{(\omega
+i/\tau+i\omega_c^2/\gamma)}{(\omega  + i/\tau+ i
\omega_c^2 /\gamma)^2 - \omega_c^2} + {\cal O}(k^2), \\
\overline{\kappa}_{xy}(\omega,k) &=&
-\frac{B}{T}\frac{\frac{s^2\,T^2\,\rho  }{{\left( P + \varepsilon  \right) }^2}
- \mu \sigma_Q \,\left( \gamma\frac{\mu\rho}{P+\varepsilon}   - 2i\frac{s\,T}{P
+ \varepsilon }
 \left( \omega+ i/\tau + i\gamma\right) \right)}{
{\left(\omega+ \imag/\tau  + \imag \,\gamma\right) }^2 -
    {{{\omega }_c}}^2}\nn\\
  &&\quad \times  \left[1- 2\pi\alpha\frac{i k}{\omega} {\sigma }_Q \frac{\left(\omega +
\imag/\tau \right)
\left( \omega + i/\tau + i\gamma  + i \omega_c^2/\gamma  \right) }{
\left(\omega+ i/\tau + i\gamma \right)^2 -\omega_c^2  }\right]\nn\\
&&\quad
+2\pi\alpha\frac{k}{\omega} \frac{B\,\mu \sigma_Q^2}{T} \frac{\left( \omega
+i/\tau\right) }{ {\left(\omega+ \imag/\tau  + \imag \,\gamma\right) }^2 -
    {{{\omega }_c}}^2} +{\cal O}(k^2),\\
\kappa_{xy}(\omega,k) &=& \frac{(\varepsilon+P)}{T} \frac{\omega_c}{(\omega +i
/\tau+ i \omega_c^2/\gamma)
^2 - \omega_c^2}+
  {\cal O}(k^2).
\eea
Note that the interaction-induced damping frequency for $\kappa$ is $\omega_c^2/\gamma$,
and not $\gamma$. The former damping frequency also appears \cite{nernst} in the response functions
for $\rho_{xx}$ and $\rho_{xy}$, as can be easily checked by inverting Eqs.~(\ref{sxxf}) and (\ref{sxyf}).

As in (\ref{kappaxx}), in the presence of a magnetic field the thermal conductivity in absence of
currents is again independent of Coulomb interactions to lowest order in $k/\omega$.

\subsection{Nernst effect}
An important thermo-electric response is the Nernst effect which measures the transverse electric field $E_y$ that is established as a consequence of an applied longitudinal thermal gradient $\nabla_x T$, in the absence of electrical currents.
The ratio $e_N=E_y/(-\nabla_x T)$ is called Nernst signal and is easily obtained from the coefficients determined above as $e_N=(\sigma^{-1}\alpha)_{xy}$. It vanishes in the absence of a magnetic field and grows linearly for small fields $B$. The Nernst effect has become a popular measurement to characterize non-standard metals, such as in bismuth~\cite{BehniaBismuth} (where it was originally discovered~\cite{NernstEttinghausen}), in other semimetals~\cite{BehniaSemimetals}, in materials close to quantum critical points~\cite{BehniaQC}, as well as in superconductors
\cite{NernstSC}. All these systems share with undoped graphene the property of being far from a simple Fermi liquid. In the latter the so-called Sondheimer cancelation suppresses the Nernst signal, while it becomes very large in the systems mentioned above. By far the strongest Nernst signals (on the order of $1 {\rm mV/K}$ for fields of $1 {\rm T}$) have been observed in bismuth whose band structure exhibits close similarities with graphene with which it shares the presence of nearly massless Dirac fermions.

From the above formalism one easily obtains the full expression for the Nernst signal (at $k=0$)
\bea
e_N&=&  \frac{k_B}{e}
\frac{\varepsilon+P}{k_B T\rho}  \frac{\omega_c /\tau}{(\omega_c^2/\gamma + 1/
\tau)^2 + \omega_c^2},
\label{nernst}
\eea
where $k_B/e=86.17 \mu {\rm V/K}$ is its natural quantum unit. For small doping $\rho$, such that $\omega_c\tau \ll {\rm min}(\gamma/\omega_c,1)$, this result simplifies to
\bea
e_N(\rho\to 0)&=&  \frac{k_B}{e}
\frac{\varepsilon+P}{k_B T\rho}  \omega_c \tau=\frac{k_B}{e} \frac{\tau T}{\hbar}\frac{B}{B^*}.
\label{nernstrho0}
\eea
In relatively clean samples close to quantum criticality this may exceed the quantum unit without violating the conditions for the applicability of hydrodynamics.

In clean samples with $\omega_c\tau\gg 1$, and in the limit of large fields $\omega_c\ll \gamma$, one obtains the result
\bea
e_N&=&  \frac{k_B}{e}
\frac{\varepsilon+P}{k_B T\rho}  \frac{1}{\omega_c \tau}=\frac{k_B}{e} \frac{\Phi_{\varepsilon+P}^2}{(\rho/\rho_{\rm th})^2 (\tau T/\hbar) (B/B^*)},
\label{nernstcleanlargefiled}
\eea
which decays inversely proportional to the field strength.

\section{Conclusions}

Our main experimental predictions for the hydrodynamic cyclotron resonance in graphene
are given by the frequency-dependent conductivities in Eqs.~(\ref{sxxf}) and (\ref{sxyf}),
with the frequencies $\omega_c$ and $\gamma$ specified in Eqs.~(\ref{omegaval}) and (\ref{gammavswc}),
and the dimensionless function of the density $\Phi_{\varepsilon+P}$ estimated in Fig.~\ref{energyfig}.
This resonance occurs in a regime of weak magnetic fields where the Landau levels are not yet formed,
and the dynamics is dominated by inelastic electron-electron collisions which occur at a rate $\sim h/k_B T$.
The electronic dynamics is ``quantum critical'', and the observation of such a resonance will offer
a valuable opportunity to explore quantum criticality. As has been argued elsewhere \cite{nernst},
similar physics applies to a variety of systems in the vicinity of a quantum phase transition,
including the superconductor-insulator transition in the cuprate superconductors.

\acknowledgments
We thank S.~Hartnoll for valuable comments at the initial stages
of this work. We also thank L.~Fritz, P.~Kim, H.~Stormer, and A.~Yacoby useful discussions.
This research was supported by the NSF under grant
DMR-0537077 and by the
Swiss National Fund for Scientific Research under grant
PA002-113151 (MM).

\end{document}